\documentclass[aps,prl,twocolumn,amsmath,amssymb,groupedaddress,showpacs]{revtex4-1}
\usepackage{graphicx,color,hyperref,amsfonts}
\def\be{\begin{equation}}
\def\ee{\end{equation}}
\def\bea{\begin{eqnarray}}
\def\eea{\end{eqnarray}}

\begin{document}
\date{\today}

\title{Ensemble Inequivalence in Systems with Wave-Particle Interaction}
\author{Tarc\'isio N. Teles}
\email{tarcisio.nteles@gmail.com}
\author{Duccio Fanelli}
\author{Stefano Ruffo}
\affiliation{Dipartimento di Fisica e Astronomia, Universit\`a di Firenze and INFN, Via Sansone 1, IT-50019 Sesto Fiorentino, Italy}

\begin{abstract}
The classical wave-particle Hamiltonian is considered in its generalized version, where two modes are assumed to 
interact with the co-evolving charged particles. The equilibrium statistical mechanics solution of the model is worked 
out analytically, both in the canonical and the microcanonical ensembles. The competition between the two modes is 
shown to yield ensemble inequivalence, at variance with the standard scenario where just one wave is allowed to develop. 
As a consequence, both temperature jumps and negative specific heat can show up in the microcanonical ensemble. 
The relevance of these findings for both plasma physics and Free Electron Laser applications is discussed.
\end{abstract}


\maketitle

Systems with long-range interactions are characterized by a pair potential which decays at large distances as a power-law, with an exponent smaller or equal to space dimension. 
Examples include globular clusters~\cite{Pad90}, two-dimensional and geophysical flows~\cite{Bouchet2012}, 
vortex models~\cite{Chavanis2000}, quantum spin models~\cite{Kastner2011}, 
cold atom models~\cite{Slama2007} as well as non-neutral plasmas~\cite{Levin2008}. 
For these systems energy is non additive implying that entropy could be non concave in some 
range of values of macroscopic extensive parameters~\cite{Campa09}. This is at the origin of 
ensemble inequivalence, which in turn instigates peculiar thermodynamic properties, like negative specific heat and temperature jumps in the microcanonical ensemble. 
Ensemble inequivalence has been reported to occur in the past for gravitational systems~\cite{Thi70}, spin models~\cite{Barre2001} and two-dimensional flows~\cite{Venaille2009}. 
These are extremely interesting models per se, as well as for their theoretical implications, 
but in general they do not allow for a straightforward experimental verification of the 
predictions drawn.

A paradigmatic example of long range interacting system is represented by the so called 
wave-particle Hamiltonian, which describes the self-consistent interaction of $N$ charged 
particles with a co-evolving wave~\cite{Escan}. This is a rather general descriptive scenario, 
often invoked in different fields of investigations where the mutual coupling between particles 
and waves proves central, as e.g. in plasma physics~\cite{Ten94} and the Free Electron Laser (FEL)~\cite{Bon1990}. The equilibrium statistical mechanics solution of the celebrated 
wave-particle Hamiltonian has been so far solely carried out for the simple setting where just 
one isolated wave is allowed to exist~\cite{Yve00,Barre2004}. 
Working under this limiting, and in many respects, unrealistic  assumption, it can be shown 
that the canonical and microcanonical solutions coincide~\cite{Barre2005}. In real experiments, however, several modes are simultaneously present and interact with the bunch of co-evolving particles. Motivated by this observation and to eventually bridge the gap between theory and experiments, we here consider the generalized equilibrium solution of the reference wave-particle Hamiltonian when a second harmonic is allowed to self-consistently develop. As we shall analytically demonstrate, the insertion of an additional wave causes the ensemble inequivalence to rise, 
within a specific range of the coupling constants which yield a first order phase transition in the canonical ensemble. This conclusion is reached for a model of marked experimental value, paving the way to direct verifications and, possibly, exploitations of the general concept of ensemble inequivalence.

The starting point of our discussion is the universal wave particle Hamiltonian which can be cast in the general form:
\begin{eqnarray}
H=\sum_{i=1}^{N} \frac{p_i^2}{2}+\sum_{h=1}^{m}\delta_{h}{I}_{h}-
2\sum_{i,h}\frac{F_h}{h}\sqrt{\frac{I_h}{N}}\cos(h\theta_i-\phi_h)\,,\;\;\;\;\;
\label{eq:ham}
\end{eqnarray}
where $I_h=N|A_h|^2$ stands for the intensity of the $h$-th harmonic~\cite{Escan}. 
Here, $(\theta_i, p_i)$, the positions and {\it momenta} of the particles, are canonically conjugated variables, 
as well as  the intensity and phase of the waves, namely $(I_h,\phi_h)$. The quantity $\delta_h$ is called the detuning parameter in the context of FEL, and measures the average relative deviation from the resonance condition. 
$F_h$ are the coupling constants that determine the relative weights of the harmonics. In general, for plasma physics applications, the coupling constants depend on the details of the dielectric function and the morphology of the Langmuir waves. For the case of the FEL, they are determined by the adopted experimental setup.
In the following, and to keep a general view on the scrutinized problem, we will treat $F_h$ as constant 
control parameters and investigate the response of the systems as they get tuned. We recall that, in addition 
to the energy, the total {it momentum} of particles and waves $P=\sum_{i} p_i + \sum_h I_h$ is also a conserved 
quantity of the dynamics. In the following analysis, we will truncate the sum over the harmonics to account for 
the first two terms, $h=1$ (the first mode) and $h=m$ (the next harmonic). In plasma related applications $m=2$. 
Undulators in planar configurations allow only odd harmonics to develop and therefore $m=3$ for a Single Pass FEL.

Let us start by constructing the microcanonical equilibrium solution of the model~\footnote{The equilibrium statistical mechanics solution for the beam-plasma version of the model in the microcanonical ensemble has been considered in~\cite{Fir06}, for the simplifying setting where the strength of the coupling is neglected.}. 
To this end we first define the microcanonical measure of the system as:
\begin{equation}
\label{eq:microensemble}
d\mu_{mc}= \prod_{h}({\rm d}\phi_h {\rm d} I_h) \prod_{i=1}^N ({\rm d}\theta_i {\rm d} p_i) \,\delta(E-H)\delta(P_0-P)\,,
\end{equation}
where $E$ and $P_0$ are the total energy and {it momentum}, respectively. In the following we shall make explicit use 
of the field amplitudes $A_h$, rather than the corresponding intensities $I_h$. For convenience, we will introduce 
the energy and {it momentum} per particle, namely $\epsilon \equiv E/N $ and $\sigma \equiv P_0/N$. 
By using the Fourier representation of the delta function and performing the integral in the phase space 
coordinates $(\theta_i, p_i)$, one can rewrite the microcanonical measure as:
\begin{equation}
\label{eq:omega1}
{\rm d}{\bar \mu}_{mc}= \int_{-\infty}^{\infty} d\lambda\, e^{N {\tilde S(\epsilon,\delta,\{\phi_h\},\{A_h\},i\lambda,N)}} \prod_h {\rm d}\phi_h {\rm d}A_h\,,
\end{equation}
where:
{\small
\begin{eqnarray}
\label{eq:entropyfunctional}
&{\tilde S}(\epsilon,\delta,\{\phi_h\},\{ A_h \},i\lambda,N)=-i\,\lambda\left[\epsilon-\frac{1}{2}\left(\sum_h A_h^2\right)^2 
-\sum_{h}\delta_{h}A_h^2\right] \nonumber \\
&+ \ln \int_{-\pi}^{\pi}d\theta \exp \left[-i\lambda\, \Phi(\theta,\phi,A) \right] + 
\ln \sqrt{\frac{2\pi}{-i\lambda}} + O(\frac{1}{N})\,,
\end{eqnarray}
}and $\Phi(\theta,\phi,A)=\sum_h \frac{2F_hA_h}{h}\cos(h \theta-\phi_h).$

For mathematical convenience and without losing generality, we  operate in Eq.~\eqref{eq:entropyfunctional} a rescaling of both the energy and the detuning: $\epsilon \rightarrow \epsilon - \sigma^2/2$ and  $\delta_h \rightarrow \delta_h + \sigma$.
In order to solve the integral in \eqref{eq:omega1} we analytically continue the integrating function to the complex $\lambda$-plane and we perform a steepest descent evaluation, which is valid in the large $N$ limit. 
Following this strategy, the extremum value of $\lambda$, herafter $\lambda^\star=i \gamma$, is found to 
match the implicit equation:
\begin{eqnarray}
\epsilon=\frac{1}{2 \gamma}+\frac{1}{2}\left(\sum_{h}A_h^2\right)^2+\sum_{h}\delta_{h}A_h^2 \nonumber \\
+\frac{\int_{-\pi}^{\pi}d\theta\, \Phi(\theta,\phi,A)\, \exp \left[\gamma \, \Phi(\theta,\phi,A) \right]}{\int_{-\pi}^{\pi}d\theta\, \exp \left[\gamma \, \Phi(\theta,\phi,A) \right]}\,.
\label{gamma}
\end{eqnarray}
To proceed in the analysis we compute the total volume accessible to the system as:
{\small
\begin{equation}
\label{omegamc}
\Omega(\epsilon,\delta)\approx \int_{-\pi}^{\pi} \int_{-\infty}^{\infty} \prod_h ({\rm d} \phi_h{\rm d} A_h) \, e^{N {\tilde S}(\epsilon,\delta,\{\phi_h\},\{A_h\},\gamma,N)}\,.
\end{equation}
}For real systems, $N$ is definitely large and it is therefore legitimate to invoke in integral~\eqref{omegamc} the saddle point approximation. By dropping unimportant contributions in the large $N$ limit, one can eventually write the equilibrium entropy per particle $s(\epsilon,\delta)$ as:
\begin{equation}
\label{eq:mcentropyeq}
s(\epsilon, \delta)=\max_{\phi_h,A_h}{\tilde S}(\epsilon,\delta,\{\phi_h\},\{A_h\},\gamma)\,,
\end{equation}
where, for convenience, we have set the Boltzmann constant to one. The maximization of this entropy functional gives the following self-consistent equations:
\begin{eqnarray}
\label{eqroot1}
A_l\left(\sum_h A_h^2 + \delta_l \right)=\frac{F_l}{l}\frac{{\cal I}_{l,m}\left(2 \gamma F_1A_1\,,\,\gamma\frac{2F_mA_m}{m}\right)}{{\cal I}_{0,m}\left(2 \gamma \,F_1A_1\,,\,\gamma \frac{2F_mA_m}{m}\right)}\,,
\end{eqnarray}
where $l=1,m$, and:
\begin{equation}
{\cal I}_{n,m}(x,y)=\frac{1}{2\pi}\int_{0}^{2\pi} d\theta \cos(n \theta) e^{x\cos(\theta)+y\cos(m \theta)}\,,
\end{equation}
Notice that in the limiting case of just one wave ($m=1$), ${\cal I}_{n,m}(x,y)$ reduces to the 
standard Bessel function~\cite{Barre2004}. Few comments are mandatory at this point. In the above equations 
we have made explicit use of the condition  $\phi_h=0$, which rigorously follows the maximization procedure 
because of symmetry reasons. Moreover, the parameter $\gamma$ can be determined by combining 
Eqs.~\eqref{gamma} and \eqref{eqroot1}. A simple calculation yields the result:
\begin{equation}
\label{eqgamma}
\gamma = \left[2\epsilon + 3\left(\sum_h A_h^2\right)^2 + 2\sum_h\, \delta_h A_h^2 \right]^{-1}\,.
\end{equation}
By inspection of Eqs.~\eqref{eqroot1}, one readily finds that $A_h=0$ (which implies $I_h=0$) is always a
solution for $\delta_h > 0$. When multiple solutions exist, one should select the values of $A_h$, which correspond 
to the global {\it maxima} of the entropy~(\ref{eq:mcentropyeq}). Indeed, the intensity of the waves plays 
the role of the order parameter, enabling one to distinguish among distinct dynamical regimes of the scrutinized 
model~\footnote{In principle one can introduce an alternative quantity to investigate the equilibrium of the system. 
This is the so called bunching parameter, which measures the degree of particles' clustering. It could be however 
shown that the introduction of such and additional (global) quantity yields a redundant equation in the variational 
problem, the bunching being determined by the intensity of the wave~\cite{Barre2004,Anto2008,Car2013}.}.
By changing the strength of the coupling constants, one can go from a configuration with $I_h=0$, to a dynamical regime 
characterized by $I_h \ne 0$. As we shall illustrate in the following, both first and second order phase transition lines 
materialize in a conveniently chosen control parameter space.

\begin{figure}[!h]
\begin{center}
\vspace{0.5cm}
\includegraphics[scale=1.,width=9.cm]{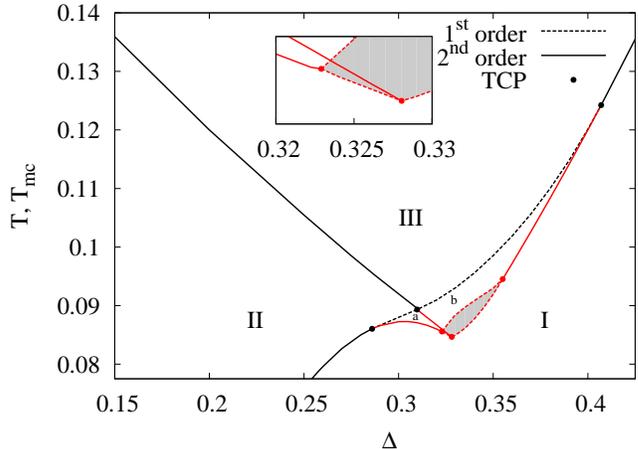}
\caption{Canonical (black online) and microcanonical (red online) phase diagram. $T$ (resp. $T_{mc}$) is the canonical
(resp. microcanonical) temperature. The dashed (resp. solid)  lines refer to the first (resp. second) order phase 
transitions in both ensembles. The filled bullets identify the canonical (black online) and microcanonical (red online) tricritical points. When the canonical phase transition is second order the ensembles give equivalent predictions. On the
other hand, inequivalence is observed in the region where the first order phase transition is predicted to occur in the
canonical ensemble (black online). In the region of inequivalence -- regions (a) and (b) -- the predicted values of 
$A_h$ differs from the corresponding $A_h^c$. The shaded area delimits the portion of the plane where temperature jumps 
occur and particularly the inset displays a zoom of the region of inequivalence. Here, and in the following, $\delta_1=\delta_m=4/3$.}
\label{fig:mcq2}
\end{center}
\end{figure}

Before turning to discuss the microcanonical phase diagram, we work out the solution of model~(\ref{eq:ham}) in the canonical ensemble. 
To this end we introduce the canonical measure:
\begin{equation}
\label{eq:canoensemble}
{\rm d}\mu_{c}=\prod_{h}({\rm d}\phi_h {\rm d} I_h) \prod_{i=1}^N ({\rm d}\theta_i {\rm d} p_i) \,  e^{-\beta H}\delta(P_0-P)\,,
\end{equation}
where $\beta=1/T$ is the inverse canonical temperature. Using again the Fourier representation of the Dirac delta function 
and repeating the basic steps of the calculations as detailed for the microcanonical analysis, one eventually obtains the 
partition function:
{\small
\be
\label{eq:partition}
Z(\beta,\delta,N)=\int_{-\pi}^{\pi}\int_{-\infty}^{\infty} \prod_h ({\rm d} \phi_h{\rm d} A_h) \,e^{-N \beta {\tilde F}(\beta,\delta,\{ \phi_h\},\{ A_h\})\}}\,,
\ee
}
where:
\bea
&{\tilde F}(\beta,\delta,\{ \phi_h\},\{ A_h\})=\sum_h \delta_hA_h^2 + \frac{\left(\sum_h A_h^2\right)^2}{2}\nonumber \\
&-\frac{1}{\beta}\ln \int_{-\pi}^{\pi}d\theta\, \exp\left[\beta\, \Phi(\theta,\phi,A)\right]\;\;\;~.
\eea
Hence, for sufficiently  large $N$, the equilibrium free energy per particle $f(\beta,\delta)$ reads: 
\be
f(\beta,\delta)=\min_{\phi_h,A_h}{\tilde F}(\beta,\delta,\{ \phi_h\},\{ A_h\})
\ee
The {\it minima} of the free energy functional ${\tilde F}(\beta,\delta,\{ \phi_h\},\{ A_h\})$ identify 
the equilibrium configurations and are determined by numerically solving the self-consistent equations:
\begin{eqnarray}
\label{eqcroot1}
A_l^{c}\left(\sum_h (A_h^{c})^2 + \delta_l \right)=\frac{F_l}{l}\frac{{\cal I}_{l,m}\left(2 \beta F_1A_1^c,\,\beta \frac{2F_mA_m^c}{m}\right)}{{\cal I}_{0,m}\left(2 \beta F_1A_1^c,\,\beta \frac{2F_mA_m^c}{m}\right)},\;\;\;\;\;\;
\end{eqnarray}
The apex $c$ has been here introduced to mark the difference with the analogue microcanonical quantities.
In the limiting case where just one wave is allowed for, the canonical solution displays a second order phase transition, 
as reported in the literature~\cite{Fir06,Escan,Yve00}, with the associated critical temperature $T_h=\frac{F_h^2}{h^2\,\delta_h}$. A question which naturally arises is whether the competition among different
modes ($F_1\ne 0$ and $F_m\ne 0$) can change the features of the transition, by turning first order the second 
order phase transition. Motivated by this working hypothesis, we here set $m=2$, a choice which
amounts to specialize on plasma physics applications. To proceed in the analysis, we will further impose the additional 
constraint $F_1+F_2=1$ which allows one to scan a finite interval of possible values of the couplings.
Denoting in particular $F_1 = \Delta$, which obviously implies $F_2=1-\Delta$, the parameter $\Delta$ 
belongs to the interval $[0,1]$. The single wave limiting solutions are respectively recovered for 
$\Delta=0$ and $\Delta=1$. The canonical phase diagram, as obtained by solving the set of self-consistent equations (\ref{eqcroot1}), is depicted in the $(T,\Delta)$  plane, see Fig.~\ref{fig:mcq2}. 

Three different phases can be identified. At large temperatures, both $I_1$ and $I_2$ are identical to zero (region III), 
as clearly expected. When the temperature gets reduced, the system experiences a phase transition towards an organized configuration. If $\Delta$ is sufficiently small, only the second wave develops (region II), while the first gets asymptotically damped.  When $\Delta$ becomes larger than a given threshold, both $I_1$ and $I_2$ are different from 
zero (region I). The transition from region II to region III (and viceversa) is second order (solid line, black online).
The border between regions I and III can be instead crossed through either a first (dashed line, black online) 
or second order (solid line, black online)  phase transition. Similar considerations apply to the adjacent regions I and II. 
Three canonical tricritical points can be identified, separating first and second order transition lines. 
These critical points are represented as bullets (black online) in Fig.~\ref{fig:mcq2}.

To compare the canonical phase diagram to its corresponding microcanonical analogue, as predicted by the theory~\eqref{eqroot1}, we introduce the microcanonical temperature. It is straightforward to see that $T_{mc} = \gamma^{-1}$, where $\gamma$ is defined by Eq.~(\ref{eqgamma}). From a direct inspection of Fig.~\ref{fig:mcq2}, one clearly realizes that the 
canonical and the microcanonical calculations yield different solutions, in the region where the competition of 
the waves is relevant at $\Delta \simeq 1/3$. 

\begin{figure}[!h]
\begin{center}
\vspace{0.5cm}
\includegraphics[scale=1.,width=9.25cm]{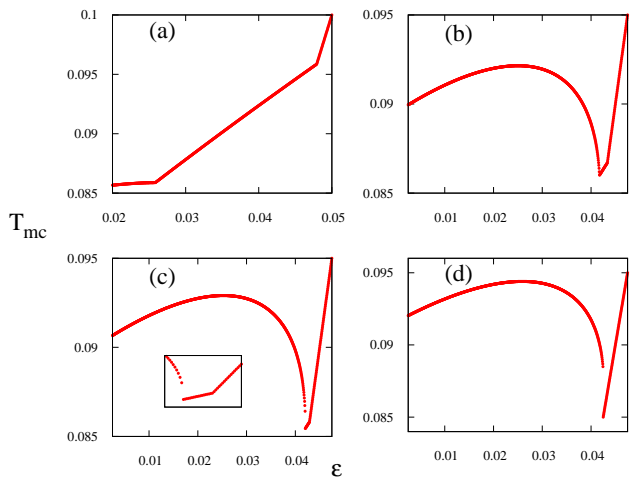}
\caption{Microcanonical caloric curves $T_{mc}$ vs. $\epsilon$ are plotted for different choices of $\Delta$.
In panel (a) $(\Delta=0.285)$ both second-order phase transitions occur in a region where ensembles are equivalent. 
In the panel (b) $(\Delta=0.32)$ both second-order phase transitions take place inside the region of ensemble inequivalence. 
The first phase transition displayed in panel (c) $(\Delta=0.325)$ is discontinuous and yields a temperature jump. The subsequent phase transition is continuous. In panel (d) $(\Delta=0.33)$ just one first-order phase transition takes place from region I to region III. Panels (b), (c), (d) shows the existence of negative specific heat in the microcanonical ensemble.}
\label{fig:mct}
\end{center}
\end{figure}

Several interesting observations can be made to appreciate the physical consequences of the predicted inequivalence. 
First, we notice that three tricritical point are also found in the microcanonical setting. Imagine to assign a 
value of $\Delta$ which is smaller than the value for which the leftmost microcanonical tricritical point is found, 
but larger than the value that identifies the location of the intermediate canonical tricritical point. 
When starting in the microcanonical region I and moving vertically (keeping $\Delta$ fixed) in the microcanonical 
phase diagram, from small to large $T_{mc}$, one induces a smooth transition towards region II. Interestingly, and 
at odd with what happens in the canonical setting for an identical choice of $\Delta$, region III can also be 
reached by acting on the temperature $T_{mc}$. In other words, while in the canonical ensemble, for dedicated choices 
of $\Delta$, only two adjacent regions can be explored when adjusting the temperature, in the microcanonical 
ensemble it is in principle possible to go through all three, macroscopically distinct, states by continuously 
changing the energy, hence the microcanonical temperature, of the system. This is a striking conclusion:  canonical 
and microcanonical predictions are qualitatively different, an observation that opens up the perspective to challenge experimentally the statistical mechanics predictions of ensemble inequivalence. In panel (b) of Fig.~\ref{fig:mct} the corresponding caloric curve, $T_{mc}$ vs. $\epsilon$ is reported. The cascade of consecutive transitions yields the 
two cusp like points, which in turn imply a punctual discontinuity in the derivative of $T_{mc}$ as function of the energy. Importantly, before the first transition line, the system displays negative specific heat in the microcanonical ensemble, 
as $T_{mc}$ decreases when $\epsilon$ gets increased. 
This is the typical signature of ensemble inequivalence: negative specific heat can be measured in the microcanonical ensemble in the region of the parameter that falls in between the canonical (black online) and microcanonical (red online) transition lines in  Fig.~\ref{fig:mcq2}. For comparison, in panel (a) of Fig.~\ref{fig:mct}, the caloric curve is plotted for a value of $\Delta$ outside the region where the two ensembles return the same predictions ($\Delta=0.285$). 
The two cusps bear the imprint of the two consecutive transitions as experienced by the system, but in this case the microcanonical specific heat is positive.

Particularly interesting is the region enclosed by the three microcanonical tricritical points and depicted with a shaded area in Fig.~\ref{fig:mcq2}. Here, temperature jumps occur: this is clearly testified in panel (d) of Fig.~\ref{fig:mct}, 
where the finite gap opens in $T_{mc}$ when progressively tuning the energy amount. Note also that, by increasing the energy, one experiences a negative temperature jump, from a hotter to a cooler state. This reflects in fact the existence of a negative specific heat region which anticipates the transition. Even more interesting is the situation schematized in panel (c) of Fig.~\ref{fig:mct}: $\Delta$ is now set to a value that falls in between the two leftmost microcanonical tricritical points, see also the zoom in the inset of Fig.~\ref{fig:mcq2}. Now the system experiences two nested transitions, respectively of first and second order, and presents both temperature jumps and negative specific heat. In the magnified image of Fig.~\ref{fig:mcq2} the microcanonical second order and first order lines seem to cross in one point of the parameters space. Clearly, this is just an apparent intersection, due to the fact that we project the microcanonical phase diagram on the plane $(\Delta, T_{mc})$ so to favor a comparison with the canonical predictions. The crossing is resolved if one operates in the more appropriate parameter space $(\Delta, \epsilon)$, when dealing with the microcanonical solution~\cite{Teles2012}.

As concerns the case with $m=3$, which, we recall, proves adequate to describe the dynamics of a 
Single Pass Free Electron Laser, one obtains a different phase diagram. The details of the diagram are not given here but again the canonical and microcanonical predictions differ. This case will be explored in a future work.

Summing up, we have here considered a straightforward generalization of the classical wave-particles Hamiltonian. 
We have in particular analyzed the general case where multiple modes are simultaneously present, and then specialized, 
for pedagogical reasons, on the simple case study where just two co-evolving waves can get unstable. 
The competition between different modes drives the emergence of ensemble inequivalence, a feature which is instead 
lacking in the limit where just one wave is solely allowed for. A plethora of phenomena, including negative specific 
heat and temperature jumps in the microcanonical ensembles, is revealed depending on the strengths of the couplings, 
assumed as tunable control parameters. Given the general interest of the investigated model, bearing in mind its 
applications to the study of both plasma instabilities and FEL dynamics, we believe that our results could constitute 
an important step forward in the search of an experimental verification of ensemble inequivalence. It should be recalled however that wave-particle systems can be trapped in long-lasting out-of-equilibrium 
states~\cite{Yama2004, Teles2010, Figuei2013, Levin2014, Anto2008, Car2013}. 
These latter might, at least in principle, coincide with the experimentally accessible regimes~\cite{Yve00}, 
so making it intricate to eventually sample the equilibrium configurations via direct measurements.
This work was supported by CNPq-Brazil.


\end{document}